\documentstyle[epsf]{article}
\def\pri{^{\, \prime}}

\def\prd#1{{\em Phys.~Rev.}~{\bf D#1}\ }

\def\prl#1{{\em Phys.~Rev.~Lett.}~{\bf #1}\ }
\def\plett#1{{\em Phys.~Lett.}~{\bf #1B}\ }
\def\np#1{{\em Nucl.~Phys.}~{\bf B#1}\ }
\def\deg{\ifmmode{^{\circ}}\else ${^{\circ}}$\fi}
\def\ni#1{\noindent$(#1)\quad$}

\def\gsim{\,\raisebox{-0.13cm}{$\stackrel{\textstyle>}{\textstyle\sim}$}\,}

\def\bi{\begin{itemize}}
\def\ei{\end{itemize}}
\def\ed{\end{document}}
\def\be{\begin{equation}}
\def\ee{\end{equation}}
\def\bea{\begin{eqnarray}}
\def\eea{\end{eqnarray}}
\def\beas{\begin{eqnarray*}}
\def\eeas{\end{eqnarray*}}
\def\req#1{(\ref{eq:#1})}
\def\eq#1{Eq.~(\ref{eq:#1})}
\def\labeq#1{\label{eq:#1}}
\def\vev#1{\left<{#1}\right>}
\def\tfrac#1#2{{\textstyle\frac{#1}{#2}}}

\def\thalf{\tfrac{1}{2}}
\def\tquarter{\tfrac{1}{4}}

\def\ev{\ \mbox{eV}}
\def\gev{\ \mbox{GeV}}

\def\mev{\ \mbox{MeV}}

\def\gsim{\raisebox{-0.5ex}{$\stackrel{>}{\sim}$}}

\def\eb{\end{thebibliography}}

\def\nn{\nonumber}
\def\nns{\nn\\[.1in]}

\def\labeq#1{\label{eq:#1}}
\def\req#1{(\ref{eq:#1})}
\def\eq#1{Eq.~(\ref{eq:#1})}
\def\bb{\bibitem}
\def\pd#1#2{\frac{\partial #1}{\partial #2}}

\def\order#1{\ifmmode {{\cal O}(#1)}\else ${\cal O}(#1)$\fi}
\def\rl{\rho_{\Lambda}}
\def\rlq{\ifmmode\rl^{\; 1/4}\else $\rl^{\; 1/4}$\fi}
\def\nf{\ifmmode N_f\else $N_f\ $\fi}
\def\tnf{2\nf}
\def\sut{\ifmmode SU(2)\pri\else $SU(2)\pri$\fi}
\def\lsut{\ifmmode \Lambda_{\sut}\else $\Lambda_{\sut}$\fi}
\def\agp{\alpha_{\rm GUT}\pri}
\def\tpt{\frac{T\pri}{T}}
\def\fct{F_c/T}
\def\tc{T_c}
\def\p{\Phi}
\def\pd{\p^{\dagger}}
\def\pdp{\pd\p}
\def\tr{\mbox{Tr}}
\def\lm{\lambda}
\def\lw{\lm_1}
\def\lt{\lm_2}
\def\lb{{\bar\lm}}
\def\lbt{\lb_2}
\def\ppl{\phi_+}
\def\pp{\phi\pri}
\def\rp{r\pri}
\def\ttc{T/\tc}
\def\vbar{{\overline V}}
\begin{document}
\begin{titlepage}
\begin{flushright}
{\sl NUB-3208/00-Th}\\
hep-ph/0003197
\end{flushright}
\vskip 0.5in

\begin{center}
{\Large\bf Proposal for a Constant Cosmological Constant}\\ [.5in] {Haim
Goldberg}\\ [.1in]
{\it Department of Physics\\ Northeastern University\\ Boston, MA 02115, USA}\\
\end{center}
\vskip 0.4in

\begin{abstract}
It is proposed that the apparent positive acceleration of the cosmological
scale factor is due to the vacuum energy of an incomplete chiral phase
transition in a hidden $SU(2)$ sector. Constraints from primordial
nucleosynthesis imply that the present metastable phase is in a substantially
supercooled state. It is argued that  massless chiral condensates can
substantially enhance the possibility of supercooling, and a linear sigma
model exhibiting scale invariance broken only at the quantum level is shown to
accommodate the required supercooling with a reasonable choice of
quartic couplings. The extensive supercooling can in principle be
confirmed or rejected on the basis of interface tension measurements
in lattice simulations with dynamical fermions.
\end{abstract}
\end{titlepage}
\setcounter{page}{2}

\section{Introduction}
There is accumulating evidence that the expansion rate of the universe is
greater now than in the past \cite{perlmutter}. The data, taken alone or in
conjunction with constraints from large scale structure \cite{turnperlwhite},
are compatible (in a flat universe)
with a contribution $\Omega_{\Lambda}\simeq 0.7$ from a
cosmological constant
$\Lambda $ or its equivalent. Cluster abundance estimates for
the matter fraction  $\Omega_{matter}$ of the critical density, combined with
an analysis \cite{dodelson} of CMB observations in the Doppler peak region
(which
support the inflation prediction  $\Omega_{tot}=1$)
are also compatible with such a contribution to the ``dark energy''.
The resulting energy density is given by
\be
\rl=(2.2\times 10^{-3}\ {\rm eV})^4\cdot
(\Omega_{\Lambda}/0.7) \cdot(h/0.65)^2\ \ ,
\labeq{rholambda}
\ee
where $h=$ present Hubble constant $H_0$ in units of 100 km/sec/Mpc. The
introduction of a non-zero value for $\rl$ or its equivalent presents an
important challenge to particle physicists and cosmologists. Even if one
concedes ignorance and simply accepts as premise that the universe is relaxing
to a state with $\rl=0$ \cite{witten},  there  still remains the perplexing
question of the origin of a mass scale $\rlq\sim 10^{-3}$ eV which makes $\rl$
relevant in the present era. Discussion in recent years has centered on models
in which $\Lambda$ becomes time-dependent, originating in the energy density of
a scalar field $\phi$ evolving in a potential $V(\phi).$ One candidate for
$\phi$ is a pseudo-Nambu-Goldstone boson (axion)  in a harmonic potential
associated with the breaking of a $U(1)$ symmetry to $Z_N$ \cite{hillross}. The
cosmological consequences of this scenario depend on both the normalization of
the potential $M^4$ and the scale $f$ at which the $U(1)$ symmetry is realized
in the Goldstone mode, and there have been studies \cite{FriemanHillWatkins,
FriemanHillStebbinsWaga} where the parameters make the model relevant
to late-time cosmology, including recent applications to the
SNeIa results \cite{FriemanWaga,CormierHolman}. The required scale
$M\sim 10^{-3}\ev$ can be associated with a
neutrino mass \cite{FriemanHillWatkins} or
with the confining scale of a hidden gauge theory \cite{Starkman},  $\phi$ being
the axion. In quintessence models  \cite{peebles,steinhardt}, $\phi$  evolves
so that the late-time behavior of the dark energy density $\rho_{\phi}$ is
largely independent of initial conditions. What remains  to be tuned by hand is
the parameter in the potential which allows for a positive acceleration of the
scale parameter during the SNeIa era relevant to the observations of Ref.
\cite{perlmutter}. The origin of the field $\phi,$ the form of its potential,
and its relation to other physics remain to be explained \cite{albrecht}.

In this paper, I would like to propose that the dark energy $\rl$ is not
evolving, but is the false vacuum energy associated with an incomplete chiral
phase transition in a hidden \sut\ gauge theory with strong scale $\sim\rlq.$
It will be seen that in the context of modern D-brane physics, the scale
$\rlq$ for the vacuum energy can be accommodated in a
natural manner in a supersymmetric GUT theory. Instead, the central problem will
be to explain how  a low temperature
$T_{hidden}<T_{\rm CMB}\simeq \tfrac{1}{10} \rlq$ can be sustained
for the quark-gluon
phase of the (supercooled) plasma of the hidden sector. Such a high degree of
supercooling can potentially be tested in lattice simulations. In the present work,
it will strongly constrain the effective field theory.
The model is described in the
next section, and some possible advantages as an alternative to the scalar field
scenario are mentioned in the concluding section.

\section{The Model}

I will consider a hidden unbroken \sut\  Yang-Mills theory whose low energy
matter content is \nf Dirac fermions (2\nf Weyl fermions) with vector-like
coupling to the gauge field. (For now, hidden sector quantities will be denoted
by a prime.) Except for gravitational interactions, this
\sut\  is completely decoupled from all standard model fields. The choice of
\sut\ is doubly motivated: (1) the running of the gauge coupling is slow,
so that the scale  \lsut\ can be pushed to values approximating
\rlq\ \cite{FriemanHillWatkins} (2) there will be fewer  massless degrees of freedom
to perturb the successful scenario of Big Bang Nucleosynthesis (BBN), a
critical requirement of the model. The hidden nature of the \sut\ will be
discussed when the coupling requirement at GUT energies is obtained.

Since the theory will be considered to have evolved from GUT energies, the
supersymmetric version becomes relevant.
The matter content then consists
solely of
2\nf chiral \sut\ doublet superfields $Q_i (i=1\ldots 2\nf).$ In order to
preserve the low energy chiral phase transition, \sut\ singlet, flavor
antisymmetric  mass terms $\sim Q_iQ_j-Q_jQ_i$ must be prohibited by  a
discrete symmetry $(R$-invariance or $Z_N, N>2$ symmetry). Next,  the
1-loop RG equation relates the gauge coupling at GUT
$\agp$ and the strong coupling scale \lsut:
\bea
\frac{2\pi}{\agp}&=&b_2^{\rm SUSY}\ln\left( \frac{M_{\rm GUT}}{1\ {\rm
TeV}}\right) + b_2^{\rm non-SUSY} \ln\left(\frac{1\ {\rm TeV}}{\lsut}\right)\nns
b_2^{\rm SUSY}&=&6-\nf\nn\\
b_2^{\rm non-SUSY}&=&\tfrac{22}{3}-\tfrac{2}{3}\nf\ \ ,
\labeq{rg}
\eea
so that for $M_{\rm GUT}=2\times 10^{16}\ {\rm GeV},$ $\lsut=\rlq=2.2\times
10^{-3}\ {\rm eV}$ one obtains

\be \agp {}^{-1}= 68.6-8.46\ \nf\ \ .
\labeq{algp}\ee

The major premise of the model is the existence of a false vacuum at present. A
first order phase transition driven by fluctuations of chiral condensates
is strongly indicated by
theoretical arguments \cite{amit,wilczek,kogut}, most cleanly for
$\nf\ge 4.$ From \eq{algp}, it would seem that unification with standard model
couplings ($\alpha_{\rm GUT}\simeq\tfrac{1}{25}$) can be achieved with a choice
$\nf=5;$ however,  this would elevate the number of Goldstone
degrees of freedom below the critical temperature $(=2\nf^2-\nf-1)$
\cite{peskin} to a
value larger than the effective number of quark-gluon degrees of freedom
$(=7\nf+6),$ presumably vitiating the first order phase transition. Thus I
choose $\nf=4,$ and
\eq{algp} gives
\be
\agp \simeq \tfrac{1}{35}\ne \alpha_{\rm GUT} \ \ .
\labeq{algpp}
\ee
Such a disparity in GUT-scale gauge couplings is not difficult to accommodate in
current formulations of string theory, with gauge fields residing in open
strings tied to D-branes. If, for example, the standard model gauge group lives
on a 5-brane and the hidden \sut\  on another 5-brane (orthogonal with
respect to the
compactified 2-tori) \cite{shiutye}, the ratio of the gauge couplings would be
inversely proportional to the volumes of the 2-tori:

\be \frac{\agp}{\alpha_{\rm GUT}}=\frac{v_2}{v_2\pri}\ \ .
\labeq{tori}
\ee
Thus, a 20\% difference in toroidal moduli could account for the disparity in
the $\alpha$'s.

\subsection*{Temperature Constraints}
Having established a model, one can quickly ascertain the constraints which
follow from BBN. The hidden sector energy density $\rho\pri$ of the \sut\
gauge fields and 2\nf\  Weyl doublets, relative to a single species of
left-handed neutrino is given by
\be
\left.\frac{\rho\pri}{\rho_{\nu_e}}\right|_{BBN}=
\left(\frac{7\nf+6}{(7/4)}\right)\left.\left(\tpt\right)^4\right|_{BBN}\ \ .
\labeq{bbna}
\ee
Requiring this ratio to be $\le 0.3$ \cite{bbn} implies (for $\nf=4)$
\be
\left.\tpt\right|_{BBN}\le 0.353 \ \ .
\labeq{bbnb}
\ee
Much of  this can  be accounted for through reheat processes in the visible
sector. Assuming no reheat for the hidden sector for energies below the
electroweak scale, one finds
\bea
\left.\tpt\right|_{BBN}&=&\left(\frac{g^*(BBN)}{g^*(>EW)}\right)^{1/3}\left.
\tpt\right|_{>EW}\\
&=&\left(\frac{10.75}{106.75}\right)^{1/3}\left.
\tpt\right|_{>EW}\ =\ 0.465\left.
\tpt\right|_{>EW}\ \ ,
\labeq{bbnc}
\eea
where $>EW$ denotes temperatures above the electroweak scale. In order to
comply with the BBN requirement (Eq. 6) an additional suppression of $T\pri/T$
above the electroweak scale by a factor of 0.76 is required. Here one can
invoke an asymmetric post-inflation reheat into visible and hidden sector
quanta \cite{mohapatra}. In the slow reheat scenario \cite{abbott},
$T_{reheat}$ is proportional to the coupling of the inflaton to the quanta
\cite{lindebook}, so
that a ratio of couplings of the same order as the ratio of the $\alpha$'s
\req{algpp} could provide the desired additional suppression. This reheat
asymmetry could have the same origin as the $\alpha$ asymmetry, if the inflaton
originates in the modular sector. Asymmetric reheating also obtains in the
case of parametric resonance \cite{linde} because of the asymmetric coupling
\cite{mohapatra}.

Because of $e^+e^-$ annihilation, the present $T\pri$ is further depressed
relative to the present CMB temperature by the same
$(4/11)^{1/3}$ factor as with neutrinos. Thus, together with \req{bbnb},
one obtains
$(T\pri_{\rm now}/T_{\rm CMB})\le
0.251.$ Since $T_{\rm CMB}=2.35\times 10^{-4}\ {\rm eV} = 0.11\ \rlq,$
one finds
\be
T\pri_{\rm now}/\rlq \le 0.028\ \ .
\labeq{tnow}
\ee
This suggests a great deal of supercooling, which needs to be accommodated.
The calculations in this work will require the ratio $\ttc$ ($\tc$
is the critical
temperature\footnote{From here on, all temperatures will be understood to be
hidden sector temperatures, and the primes will be omitted.}) which in turn
requires knowledge of the ratio $\rlq/T_c.$ This will be calculable in the
effective field theory to be discussed.

\subsection*{Supercooling}
In the standard formulation of first order phase transitions via critical
bubble formation \cite{landau,alcock} the condition for {\em failure} to
complete a phase transition in the expanding universe at (hidden) temperature
$T$ is \cite{guth}
\be (T/H(T))^4\ e^{-\fct}< 1\ \ ,
\labeq{crita}
\ee
where $F_c$ is the free energy of a critical bubble, and $H(T)$ is the Hubble
constant at temperature $T.$ With $H_0\simeq 2.2\times 10^{-33}\ h$ eV,
$T=0.28\ T_{\rm CMB},$ one obtains the condition for failure to nucleate
\be \fct> 260\ \ .
\labeq{critb}
\ee
In the thin wall approximation, the bubble has  a well-defined surface tension
$\sigma,$ and the picture is consistent only for small supercooling below
$\tc.$ The bubble action is given by \cite{landau,alcock}
\be \frac{F_c}{T}=\frac{16\pi}{3}\ \frac{\sigma^3}{L^2\eta^2\tc}\ \ ,\labeq{fcta}
\ee
where $L$ is the latent heat and $\eta=(\tc-T)/\tc.$ Thus, a failure to
nucleate via thin-walled bubbles requires a large surface tension,
$\sigma/\tc^3\ \gsim\  1.$ In lattice studies of quenched QCD \cite{iwasaki},
the interface tension between confined and deconfined phases is small:
$\sigma/\tc^3\simeq 0.1.$ However, a simple
calculation \cite{pokrovskii} based on the MIT bag model \cite{BagModel} suggests
that the picture can change  drastically in
the presence of chiral condensates: in that case,
\be \sigma=-\tquarter\sum_{i=1}^{N_f} \vev{\bar q_iq_i}\ \ .
\labeq{pokrov}\ee
which for QCD $(\tc\simeq 150\ \mev, \vev{\bar q_iq_i}
\simeq -(240\ \mev)^3$ per flavor) would give a large surface tension,
$\sigma\simeq 4\tc^3,$ possibly invalidating the thin-wall
approximation.\footnote{This would give an average distance
between nucleation sites of
about 10 m \cite{kajantie}, perhaps marginally affecting
primordial light element abundances \cite{alcocka}.} A
full analysis of the unquenched QCD situation is probably best carried in the
framework of a mean field theory \cite{campbell}. I will proceed in the context
of such a theory to see what constraints are imposed on the \sut\ model in order
to attain the desired metastability (\eq{critb}) until the present era.

\section{Linear Sigma Model for \sut\ with \nf Flavors.}
The symmetry breaking pattern of color \sut\  with \nf flavors has long been
known \cite{peskin}, and a  linear sigma model for this case has recently been
examined \cite{wirstam}. Such a  model will serve conveniently to study the
chiral phase transition. The meson and diquark baryon fields
are contained in the $\tnf\times \tnf$ antisymmetric matrix $\p_{ij}=-\p_{ji},$
with chiral symmetry breaking occurring in $Sp(\tnf)$ direction \cite{peskin}
compatible with the
Vafa-Witten theorem \cite{vafa,peskin,koguta}
\be
\left<\p\right>_0=\frac{\phi_0}{\sqrt{2(\tnf)}}\left(
\begin{array}{ccc}
{\bf 0}& {\bf 1}\\
{\bf -1} & {\bf 0}
\end{array}\right)\ee
where $\bf 0$ and $\bf 1$ are $\nf\times \nf$ matrices. The lagrangian is
\be {\cal L}=\tr\ \partial^{\mu}\p\partial_{\mu}\p -
m^2\ \tr\ \pdp-\lw\left(\tr\ \pdp\right)^2-\lt\tr\ \pdp\pdp\ \ .
\labeq{lagr}
\ee
I have omitted a term $\propto \mbox{Pf}(\p)+\mbox{Pf}(\pd)$ arising from the
axial anomaly. Since I will be working with $\nf=4,$ this additional operator
quartic in the fields will not qualitatively change  the discussion which follows.
Stability in all field directions requires $\lt\ge 0,\ \lw+\lt/\tnf\ge 0.$

Specializing now to the field $\phi$ in the direction of the vev, one obtains
\bea
{\cal L}_0& = &\thalf(\partial_{\mu}\phi)^2 -V(\phi)\ \ ,\nns
V(\phi)&=&\thalf\ m^2\phi^2+\tquarter\ \lb\phi^4\ \ ,\nns
\lb&=&\lw+\lbt,\ \ \lbt=\lt/\tnf \ \ .
\labeq{ello}
\eea
With finite temperature corrections (restricted for
simplicity to the $m^2$ term) and the introduction of a running quartic
coupling,
one obtains the effective potential
\be
V(\phi,T)= \thalf\ m^2(T)\phi^2 + \tquarter\ \lb(t)\phi^4\ \ ,
\labeq{veff}
\ee
where $t=\ln(\phi/\phi_0)$, and $m^2$ has the standard $T$-corrected form
\be
m^2(T)=A(T^2-T_0^2)\ \ .
\labeq{msq}
\ee
In the model described, $A$ can be calculated, and I find at one loop
\be
A=\tfrac{1}{12}\left[(\tnf(\tnf-1)+2) \lb + (6\nf(\tnf-1)-2)\lbt\right]\ \ .
\labeq{a}
\ee
The $\epsilon$ expansion analysis of the model described by \req{lagr} shows
that it allows a first order phase transition through a Coleman-Weinberg
mechanism  at $T=T_0,$ when $m^2(T)=0$ \cite{wirstam}.
However,
it will  shortly be apparent that the large supercooling will
require that  $T_0^2\ll T_c^2.$  This (approximate) conformal invariance at tree level in the chiral lagrangian (to be discussed below)  in turn suggests
that
chiral symmetry breaking at {\em zero temperature}  in this model
also proceeds through radiative
corrections (Coleman-Weinberg) \cite{coleman}. Thus, to lowest order (see
\eq{yamag} below)
\be
\lb(t)=-\lm(1-4t)\ \ ,\quad \lm\equiv -\lb(0)\ \ .
\labeq{lamb}
\ee
where $t=0$ is defined by the minimum of the second term in
\req{veff}.

The vacuum at $\phi=0$ described by the potential \req{veff}
becomes metastable at a temperature $T_c$ determined by
requiring simultaneously
\be
V\pri(\ppl)=V(\ppl)=0\ \ ,
\labeq{vv}
\ee
at some field value $\ppl.$ A short algebraic
exercise with Eqs. \req{veff}, \req{msq}, \req{lamb}, and \req{vv}
determines $\tc:$
\be
m(\tc)=\sqrt{A(\tc^2-T_0^2)}=\sqrt{\lm}\phi_0e^{-1/4}\ \ .
\labeq{tca}
\ee
\subsection*{Suppression of $m^2$.}
For $T_0<T<\tc$ there is a barrier between the false and true vacua. But the
extreme supercooling requirement indicated in \eq{tnow} will be seen to impose
a large hierarchy, $T_0\ll\tc$ (it will turn out that $T_0\le 0.124 T_c.)$
There is no obvious argument to justify this hierarchy --- it is simply
required in this model for compatibility with the supercooling requirement.
Nevertheless, two comments may be made: (1) A quantitative comparison with the
linear $SU(3)\times SU(3)$ sigma model (whose dynamics differs through the
presence of the cubic term) is perhaps instructive. In that case, the
coefficient $A$ is obtained as a sum over the massive mesons \cite{dolan},
$A=\tfrac{1}{12}\sum_{i}M_i^2/v^2,$ where $v\simeq f_{\pi}.$ With $M_i\simeq 1$
GeV,  one estimates  $A\simeq 83.$ Since $T_0=\sqrt{-\mu^2/A},$ where
$\mu^2\simeq
-0.15\ \gev^2$ \cite{haymaker} is the temperature-independent (negative) mass
parameter, we find $T_0\simeq 43$ MeV $\simeq 0.24 \tc.$ This may provide a
normative expectation for $T_0/\tc.$ (2) In $SU(N)$ gauge theories with $N_f$
massless fermion flavors, a {\em continuous} transition to an approximately
conformal phase (which would imply $m^2=0)$ is suggested at some value of
$N_f/N<11/2\ \cite{appelquist}.$ Some analytic studies
\cite{appelquist,schechter} indicate $N_f/N\approx 4$ as a critical value,
while a QCD lattice study \cite{mawhinney} show hints of a suppression of the
chiral condensate (expected in the transition to the conformal phase
\cite{appelquist}) for a smaller value, $N_f/N=\frac{4}{3}$ ($SU(3)$ with 4
flavors). Perhaps the ratio in the present case $(N_f/N=2)$ is sufficiently
large to significantly suppress the zero temperature $m^2$ term in the
effective lagrangian
--- a lattice study of chiral symmetry breaking in
$SU(2)$ with $N_f=4$ could in principle shed light on
this question. At any rate, at this juncture I accept the hierarchy
$T_0\ll\tc,$ and simplify matters even more by setting
\be
T_0=0
\labeq{teqz}
\ee
in \eq{msq}. In such a model, with {\em two} coupling constants, the chiral
invariance is broken at $T=0$ in the Coleman-Weinberg manner
\cite{coleman,paterson}. For $T\ne 0,$ the transition becomes first order, as
described in the previous section.

\subsection*{Critical Bubbles.}
For $T\ne 0$ it will prove convenient to rescale $\phi=m(T)\pp/2\sqrt{\lambda},$
so that using Eqs. \req{veff}, \req{lamb} and  \req{tca} one may write
\bea
V &\equiv& \frac{m^4(T)}{4\lm}\ \vbar\nns
\vbar &=& \thalf{\pp}^{\ 2}+\tquarter{\pp}^{\ 4}\
\left[ \ln((m(T)/m(T_c))\pp)/2-\thalf\right]\nns
&=&\thalf{\pp}^{\ 2}+\tquarter{\pp}^{\ 4}\
\left[ \ln((\ttc)\pp/2)-\thalf\right]\ \
\labeq{barv}
\eea
for $T_0=0.$

The O(3) symmetric free energy for a critical bubble formed at temperature
$T$ is given by
\cite{colemanb}
\bea
F_c&=&4\pi\
\frac{m(T)}{4\lm}\int^{\infty}_{0}d\rp{\rp}^2\left[\thalf
\left(d\pp/d\rp\right)^2 + \vbar(\pp,\ttc)\right]\nns
&\equiv&\frac{m(T)}{4\lm}\ f(\ttc)
\labeq{fctb}
\eea
where $\rp=m(T)r.$ The field $\pp$ is the solution to
\be
\frac{d^2\pp}{d{\rp}^2}+\frac{2}{\rp}\ \frac{d\pp}{d\rp} = \frac{\partial
\vbar}{\partial \pp}\ \ .
\labeq{eqmotion}
\ee
subject to $\left.d\pp/d\rp\right|_{\rp=0}=0,\ \pp(\infty)=0.$ With the help of
Eqs. \req{msq}, \req{a} and \req{teqz} the bubble action can then be calculated
more explicitly in terms of the the quantity $f(\ttc).$
For $\nf=4,$ I obtain

\bea
\fct&=&\sqrt{A}/(4\lm)\ f(\ttc)\nns
&=&\sqrt{\tfrac{29}{6}\lb + \tfrac{83}{6}\lbt}/(4\lm)\ \ f(\ttc)\ \ .
\labeq{fctc}
\eea

At this point, I implement the condition of radiative symmetry breaking  (at
$T=0$) in the two-parameter $(\lb,\lbt)$
space. This imposes the condition \cite{yamagishi}
\be
-4\lb(0)=4\lm=\beta_{\lb}(0)\ \ .
\labeq{yamag}
\ee
{}From the one-loop RG equations \cite{wirstam}, for
$\lbt(0)\gg\left|\lb(0)\right|,$ but
$\lbt(0)$ still perturbative (this
will be justified {\em a posteriori}), \eq{yamag} gives
\be
\lbt(0)= \left(4\pi/\sqrt{4\nf^2-2\nf-2}\ \right)\sqrt{-\lb(0)}=(4\pi/\sqrt{54})\sqrt{\lm}\
\ \ ,
\labeq{lbtlm}
\ee
and hence in the same approximation
\be
A\simeq \tfrac{83}{6}\tfrac{4\pi}{\sqrt{54}}\ \sqrt{\lm}\ \ .
\labeq{aa}
\ee
Thus the bubble action can be written entirely in terms of the coupling
constant $\lm=-\lb(0):$
\bea
\fct& =&\sqrt{\tfrac{4\pi}{\sqrt{54}}\tfrac{83}{6}\sqrt{\lm}}/(4\lm)\ \
f(\ttc)\nns
&\simeq & 1.22\ f(\ttc)\ \lm^{-3/4}
\ \  .
\eea
We now require the ratio $\ttc.$ From \eq{tnow}, one needs to relate the
vacuum energy $\rho_{\Lambda}$ to $\tc.$ At $T=0,\ m(T)=0,$ Eqs. \req{veff},
\req{lamb}, \req{tca}, \req{teqz} and \req{aa} give
\bea
\rho_{\Lambda} &=& V(0)-V(\phi_0)\nns
&=&\tquarter \lm\ \phi_0^4\nns
&=&\tquarter e A^2\tc^4/\lm\nns
&=&(4.4\tc)^4\ \ .
\labeq{rhol}
\eea
Combining this with \eq{tnow}, we have the supercooling requirement
\be
\ttc\le 0.124\ \ .
\labeq{ttc}
\ee
The bubble action may now be evaluated numerically, and I find
$f(0.124)=6.61.$ Since $f(\ttc)$ is a uniformly decreasing function of $\ttc,$
I obtain (using \req{critb}) the condition for no nucleation until the present
era
\be
260\le\fct\le (1.22)(6.61)\ \lm^{-3/4} \ \ ,
\labeq{critd}
\ee
giving a bound on $\lm,$
\be
\lm\le 0.010,\ \mbox{or}\ -0.010\le\lb(0)\le 0\ \ .
\labeq{lambound}
\ee

\subsection*{Fine Tuning?}

Does the bound \req{lambound} represent a substantial fine tuning? In an
attempt to answer this question, I have presented in Figure 1 the
renormalization group flow in the $\lb-\lbt$ plane over less than or equal to
a decade
$\left[(t\le \ln(10))\rightarrow (t=0)\right]$ for those
trajectories which begin in the
stability region and satisfy the requirement \req{lambound}. It is seen
that (1) a reasonable piece of the coupling constant phase space is
available
(2) the couplings are perturbative but not particularly small  over
much of the RG flow and
(3) over some of the phase space $\lb$  reaches values of the order of the
electroweak Higgs coupling. (In the same normalization, $\lambda_{H}=
0.08\ (m_H/100\ \gev)^2.)$
\begin{figure}[ht]
\begin{center}
\epsfxsize=4.8 in
\epsfysize=4.2 in
\hfil
\epsffile{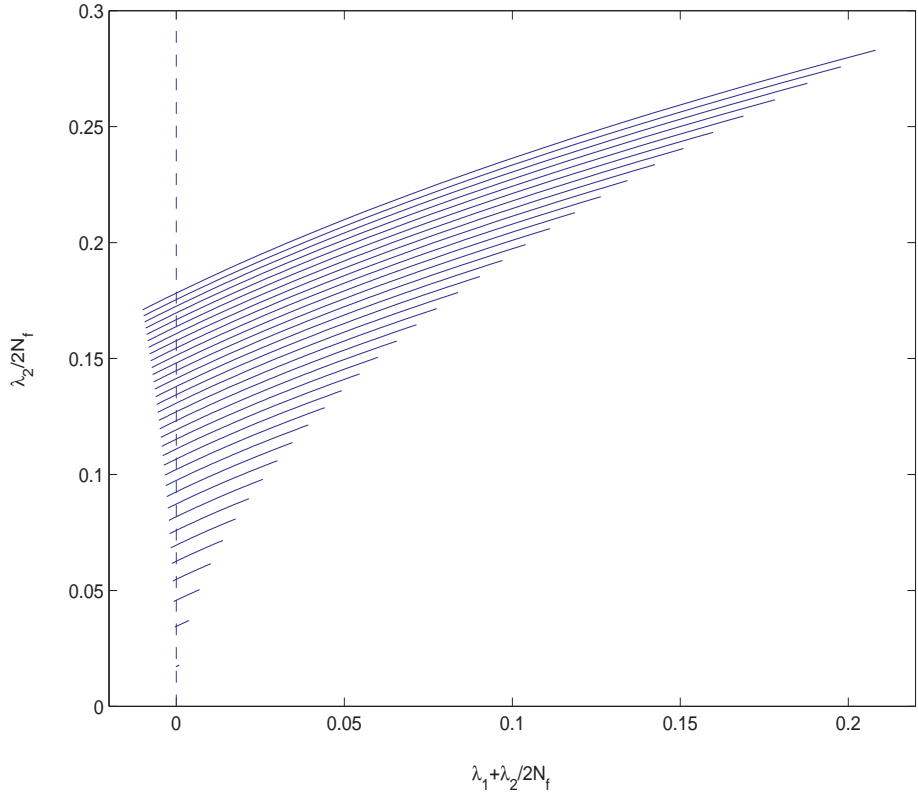}
\hfil
\caption{
Basin of attraction (for $t\le \ln(10)$) to relevant parameter space
(\eq{lambound}) at $t=0$. Region to right of dashed line is stable for all
field directions.
 }
\label{rengenf}
\end{center}
\end{figure}

A summary of  results and concluding remarks follows.

\section{Discussion and Conclusions}

\ni{1}A model has been presented  which can generate a cosmological constant
of magnitude to
account for $\Omega_{\Lambda}\simeq 0.7$ during the present era. The present
quasi-deSitter phase is driven by the false vacuum energy associated with
the supercooled
phase of an incomplete chiral phase transition in a hidden  gauge theory.
The
very small energy scale
$\rlq\simeq 2\times 10^{-3}$ eV for the vacuum energy appears as the strong
interaction scale for a hidden \sut\  whose coupling runs from GUT energies, with
coupling at GUT not quite unifying with the standard model couplings. Having
this \sut\ and the standard model fields reside on different  branes
presents a simple solution to this disparity.

\ni{2}An unchanging vacuum energy has some advantages over evolving
primordial scalar fields as an origin of the present near-deSitter phase: the
problem of protecting the tiny curvature of the potential \cite{kolda, carroll}
is circumvented, as is the necessity (in quintessence models) to control the
contribution of dark energy during nucleosynthesis \cite{albrecht}.

\ni{3}BBN considerations imply  large supercooling  in the metastable quark-gluon
plasma of the present phase. Although present lattice simulations indicate only
small supercooling in the {\em quenched} approximation of Yang-Mills theory,
theoretical considerations indicate that large supercooling is possible in the
presence of chiral condensates. A linear sigma model with two quartic
couplings was analyzed in which  all symmetry breaking takes place via
dimensional transmutation. In this model the existence of substantial
supercooling does not  require fine tuning in the coupling constant space. The
existence of a large interface energy in an $SU(2)$ theory with $\nf=4$ dynamical
quarks would provide incisive support for this model.

\ni{4}A pivotal requirement in this scenario is the near-scale invariance of the
chiral lagrangian. This was briefly discussed in the text in terms of the ratio
$N_f/N\ (=2$ in the present work). A hint
of the continuous approach to a
conformal phase may be suggested in the observed weakening of the chiral
phase transition  in a lattice study  \cite{mawhinney} of four flavor
QCD with
$N_f/N$ as small as 4/3.

\ni{5}The bound $\lm\le 0.010$ (\eq{lambound}) does not change drastically with a
tighter bound on the extra effective number of neutrinos. For example, with a
requirement
$\Delta N_{eff}\le 0.1,$ I find instead of \req{lambound} the bound
$\lm\le 0.0086.$ This would also involve a small amount of extra
reheating in the visible sector between the GUT and electroweak scales.

\ni{5}Although the phase transition discussed in this paper is long overdue, it
may not be catastrophic when it occurs. The nucleation occurs via very
thick-walled bubbles, so that the drastic shock-wave scenario depicted in the
thin-walled case \cite{colemanb} is perhaps not inevitable.

\subsection*{Acknowledgement} I would like to thank Gary Steigman for
comments and for drawing attention to an arithmetic error in \eq{bbnc}
in the original version of the manuscript. This research was supported
in part by the
National Science Foundation through Grant No. PHY-9722044.

\begin{thebibliography}{99}
\bb{perlmutter}S.J Perlmutter {\em et al.,} {\em Nature} {\bf 391} (1998) 51;
S.J. Perlmutter {\em et al.,} {\em Astrophys. J.} (1999) 517;
A.G. Riess {\em et al.,} {\em Astron. J.} {\bf 116} (1998) 1009.
\bb{turnperlwhite}S. Perlmutter, M.S. Turner and  M. White, \plett{83}(1999) 670.
\bb{dodelson}S. Dodelson and L. Knox, \prl{84}(2000) 3523.
\bb{witten}E. Witten, hep-ph/0002297
\bb{hillross}C.T. Hill and G.G. Ross, \np{311}(1988) 253; \plett{203}(1988) 125.
\bb{FriemanHillWatkins}J. Frieman, C.T. Hill and R. Watkins, \prd{46}(1992) 1226.
\bb{FriemanHillStebbinsWaga}J.A. Frieman, C.T. Hill, A. Stebbins and I. Waga,
\prl{75}(1995) 2077; see also K. Coble, S. Dodelson and J. Frieman,
\prd{55}(1997) 1851; A. Singh, \prd{52}(1995) 6700.
\bb{FriemanWaga}J. Frieman and I. Waga, \prd{57}(1998) 4642.
\bb{CormierHolman}D. Cormier and R. Holman, \prl{84}(2000) 5936.
\bb{Starkman}G. Starkman, as quoted in Ref.\cite{FriemanHillWatkins}.
\bb{peebles}P.J.E. Peebles and B. Ratra, {\em Ap. J. Lett.} {\bf 325} (1988)
L17.
\bb{steinhardt}R.R Caldwell, R. Dave and P.J. Steinhardt, \prl{80}(1998) 1582;
I. Zlatev, L. Wang and P.J. Steinhardt,
\prl{82}(1999) 896.
\bb{albrecht}For a recent discussion involving only Planck-scale physics, see
A. Albrecht and  C. Skordis, \prl{84}(2000) 2076.
\bb{amit}H.H. Iacobson and D.J. Amit, {\em Ann. Phys. (N.Y.)} {\bf 133} (1981) 57.
\bb{wilczek}R.D. Pisarski and F. Wilczek, \prd{29}(1984) 1984.
\bb{kogut}J. Kogut, \np{290}(1987) 1.
\bb{coleman}S. Coleman and E. Weinberg, \prd{7}(1973) 1888. As first shown in
this paper, two couplings are required in order that the symmetry breaking
take place in the perturbative regime.
\bb{peskin}M.E. Peskin, \np{175}(1980) 197.
\bb{shiutye}G. Shiu and S.-H. Henry Tye, \prd{58}(1998) 106007.
\bb{bbn}For recent discussion, see K.A. Olive and D. Thomas,
{\em Astropart. Phys.}{\bf 11}(1999) 403; S. Burles, K. Nollett, J. Truran and
M.S. Turner, \prl{82}(1999) 4176; E. Lisi, S. Sarkar and F. Villante,
\prd{59}(1999) 123520; K.A. Olive, G. Steigman and T.P. Walker,
{\em Phys. Rept.} {\bf 333-334} (2000) 389.
\bb{mohapatra}Z.G. Berezhiani, A.D. Dolgov and R.N. Mohapatra,
\plett{375}(1996) 26.
\bb{abbott}L. Abbott, E. Farhi and M. Wise, \plett{117}(1982) 29;
\bb{lindebook}A.D. Linde, {\em Particle Physics and Inflationary Cosmology,}
Harwood, Chur, Switzerland 1990.
\bb{linde}L. Kofman, A.D. Linde and A.A. Starobinsky, \prl{73}(1994)3195; Y.
Shtanov, J. Traschen and R. Brandenberger, \prd{51}(1995) 3438.
\bb{landau}L.D. Landau and E.M. Lifshitz, {\em Statistical Physics}, pp. 471-474
(Addison-Wesley, Reading, MA).
\bb{alcock} G.M. Fuller, G.J. Mathews and C.R. Alcock,
\prd{37}(1988) 1380.
\bb{guth}A.H. Guth and E.J. Weinberg, \prd{23}(1981) 876.
\bb{iwasaki}Y. Iwasaki, K. Kanaya, L. Karkkainen, K. Rummukainen and T. Yoshie,
\prd{49}(1994) 3540.
\bb{pokrovskii}Yu.E. Pokrovskii, {\em Sov. J. Nucl. Phys.} {\bf 50} (1989) 565.
\bb{BagModel}A. Chodos, R.L. Jaffe, K. Johnson and C.B. Thorn, \prd{10}(1974)
2599.
\bb{kajantie}C. Alcock {\em et al.,} Ref. \cite{alcock}  incorporating
correction  in K. Kajantie, L. Karkkainen and K. Rummukainen,
\np{333}(1990) 100.
\bb{alcocka}C. Alcock, G.M. Fuller, G.J. Mathews and B. Mayer,
{\em Nucl. Phys.} {\bf A498} (1989) 301.
\bb{campbell}B.A. Campbell, J. Ellis and K.A. Olive, \np{345}(1990) 57
contains a model treatment of the contribution of the gluon condensate to the
surface tension. One may note that, purely on dimensional grounds, there
is no simple analogue to \eq{pokrov} for the  pure
gauge field contribution to the surface tension in terms
of the gauge-invariant gluon condensate $\vev{G_{\mu\nu}G^{\mu\nu}}.$
\bb{wirstam}J. Wirstam, \prd{62} (2000) 045012.
\bb{vafa}C. Vafa and E. Witten, \np{234}(1984) 173.
\bb{koguta}J.B. Kogut, M.A. Stephanov and D. Toublan, \plett{464}(1999) 183.
\bb{dolan}L. Dolan and R. Jackiw, \prd{9} (1974) 3320.
\bb{haymaker}See, for instance, L.-H. Chan and R.W. Haymaker,
\prd{10}(1974) 4143.
\bb{appelquist}T. Appelquist, J. Terning, and L.C.R. Wijewardhana,
\prl{77}(1996) 1214.
\bb{schechter}F. Sannino and J. Schechter, \prd{60}(1999) 056004.
\bb{mawhinney}R.D. Mawhinney, {\em Nucl. Phys. Proc. Suppl. A} {\bf 60} (1998)
306; hep-lat/9705030; D. Chen and R.D. Mawhinney,
{\em Nucl. Phys. Proc. Suppl. } {\bf 53} (1997) 216.
\bb{paterson}This has been discussed in  detail for the QCD-based chiral
$SU(3)\times SU(3)$ model by A. J. Paterson, \np{190}(1981) 188.
\bb{colemanb}S. Coleman, \prd{10}(1977) 2929.
\bb{yamagishi}H. Yamagishi, \prd{23}(1981) 1880.
\bb{kolda}C. Kolda and D. Lyth, \plett{458}(1999) 197.
\bb{carroll}S. M. Carroll, \prl{81}(1998) 367.
\eb\ed
To obtain the amount of supercooling, I relate \rlq to the critical temperature
by approximating the entropy of the condensed phase by that of the
$2\nf^2-2\nf-2$ Goldstones (see above). This gives a latent heat $L\simeq
\rlq=T_c(s_{qg}-s_{Goldstones})=(\pi^2/90)(8\nf+7-2\nf^2)T_c^4,$ so that
\be T_c\simeq 1.0 \rlq\ \ .\labeq{tc}\ee The scenario then requires a very large
amount of supercooling:
\be T\pri/T_c\le 0.028

What are the implications of the small coupling for the meson spectrum?  As an
example, I calculate the mass of the scalar corresponding to small
oscillations in the potential
\req{veff} (with
$m(T)=0)$ and obtain
\be m_S= 12.4 \lm^{1/4}\tc = 4.9 \tc\ \ ,
\labeq{mstc}
\ee
a result not much different from that found in QCD.